# A Geometric Neutron Star Model of Repeating and Nonrepeating Fast Radio Bursts

Ze-Nan Liu[1,2], Zhao-Yang Xia[1,2], Shu-Qing Zhong[3,4], Fa-Yin Wang[1,2], and Zi-Gao Dai[4]
[1] School of Astronomy and Space Science, Nanjing University, Nanjing 210023, People's Republic of China
[2] Key Laboratory of Modern Astronomy and Astrophysics (Nanjing University), Ministry of Education, Nanjing 210023, People's Republic of China
[3] Deep Space Exploration Laboratory/Department of Astronomy, University of Science and Technology of China, Hefei 230026, People's Republic of China
[4] School of Astronomy and Space Science, University of Science and Technology of China, Hefei 230026, People's Republic of China; daizg@ustc.edu.cn
Received 2024 January 4; revised 2024 February 6; accepted 2024 February 15; published 2024 April 12

## Abstract

Fast radio bursts (FRBs) are millisecond-duration extragalactic radio transients. They fall into the categories of repeaters and apparent nonrepeaters. However, such a classification causes a lack of motivation to investigate the physical picture. Here, we propose a unified geometric model to distinguish between repeaters and apparent nonrepeaters, in which the quasi-tangential (QT) propagation effect within the magnetosphere of a neutron star is considered. In this model, apparent nonrepeaters arise from sources whose emitting region has a smaller impact angle with respect to the magnetic axis, while repeaters come from sources whose emitting region has a larger impact angle. The observational discriminant polarization properties between repeaters and apparent nonrepeaters are an important clue for verifying this unified geometric model since the polarization is sensitive to the QT propagation effect. Moreover, our model effectively explains all of the other discriminant properties, including bandwidth, duration, peak luminosity, energy, brightness temperature, time–frequency downward drifting, and repetition rate, providing compelling evidence for the magnetospheric origin of FRBs.

*Unified Astronomy Thesaurus concepts:* Radio bursts (1339); Radio transient sources (2008); Magnetars (992); Radiative processes (2055)

## 1. Introduction

Fast radio bursts (FRBs) are one of the few remaining unsolved puzzles in astrophysics (Lorimer et al. 2007). Up to now, about 1000 FRBs have been observed, and a small number of them exhibit repeating behaviors.[5] Some observational features (e.g., duration, bandwidth, spectrum) of repeaters and apparent nonrepeaters show significant distinctions (CHIME/FRB Collaboration et al. 2019). However, there is still no conclusive evidence to suggest whether they belong to one population or two populations.

Polarization is a crucial probe for studying the origins of repeaters and apparent nonrepeaters. In numerous apparently nonrepeating FRBs, a robust detection of circular polarization (CP) has been made, with certain instances revealing CP sign changes and highly diverse swings in polarization angles (PAs) across bursts (Masui et al. 2015; Ravi et al. 2016; Caleb et al. 2018; Prochaska et al. 2019; Day et al. 2020). Strikingly, a drastic variation timescale of the CP is smaller than 1 ms for apparently nonrepeating FRBs (Day et al. 2020). In contrast, repeaters demonstrate a flat CP variation within a larger timescale (Feng et al. 2022b). Despite the discrepancy in polarization observations that is evident between repeaters and apparent nonrepeaters, the question of whether these polarization disparities affect the classification of FRBs based on their repeatability remains an intriguing mystery.

Under the assumption that repeaters and apparent nonrepeaters belong to one population, the distribution differences between them for repetition rate, duration, bandwidth, and peak luminosity can be interpreted by selection effect due to the beamed emission (Connor et al. 2020; Zhong et al. 2022). However, the distribution differences for the spectral index and peak frequency cannot be explained by this simple beamed emission model. It remains uncertain whether other distinguishing properties between repeaters and apparent nonrepeaters, such as polarization and time–frequency downward drifting, can be adequately explained or not. The polarization of repeaters has been studied in the literature, for example, the CP of repeating FRBs could be caused by the intrinsic radiation mechanism (Wang et al. 2022b; Liu et al. 2023b) or the propagation effect (Dai et al. 2021; Wang et al. 2022). However, the polarization of apparent nonrepeaters has not been thoroughly discussed. The propagation effects within a pulsar magnetosphere can strongly modify radio wave intensity and polarization profiles (Wang et al. 2010). Generally, a smaller impact angle $\chi$, the angle between the line of sight (LOS) and the magnetic dipole axis, could lead to a complex CP profile and drastic variations of PAs owing to the quasi-tangential (QT) propagation effect (Wang et al. 2010). We find that the polarization of both repeaters and apparent nonrepeaters resembles the polarization profiles generated by the propagation effects within the magnetosphere. This motivates us to investigate the differences in polarization between repeating and apparently nonrepeating bursts, thus revealing their physical origins.

In this paper, we present a geometric model that enables discrimination between repeaters and apparent nonrepeaters, in which the QT effect within the magnetosphere of a neutron star is considered. Notably, nearly all observed distinguishing characteristics of repeaters and apparent nonrepeaters can be effectively explained by our model. Apparent nonrepeaters display more drastic CP and PA evolution in a shorter timescale, broader bandwidth, shorter duration, higher peak luminosity, greater energy, elevated brightness temperature, reduced drifting rate, and lower observed repetition rate compared to repeaters.

---

[5] A Transient Name Server system for newly reported FRBs (https://www.wis-tns.org; Petroff & Yaron 2020). Two more popular catalogs are https://www.herta-experiment.org/frbstats/ and http://www.frbcat.org (Petroff et al. 2016).

This paper is organized as follows. We first discuss the propagation effects within the magnetosphere of a neutron star in Section 2, and present a geometric model for both repeaters and apparent nonrepeaters in the Appendix. The explanation of observations for repeaters and apparent nonrepeaters in our model is discussed in Section 3. The results are discussed and summarized in Section 4. The convention $Q_x = Q/10^x$ in cgs units is used throughout this paper.

## 2. Propagation Effects within the Magnetosphere

Many propagation effects within the magnetosphere of a neutron star could affect the wave intensity and polarization, mainly including the resonant cyclotron absorption, wave mode coupling, and wave propagation through the QT regions (Wang et al. 2010). We briefly discuss these propagation effects as follows. Cyclotron absorption of radio emission within a pulsar magnetosphere can generate CP (Blandford & Scharlemann 1976; Wang et al. 2010). Additionally, cyclotron absorption can be applied to FRBs, as shown by Qu & Zhang (2023). They considered FRBs produced in the open field line region of a magnetar; an asymmetric distribution of the electron Lorentz factors is needed to produce a relatively high CP fraction. However, the cyclotron absorption effect only changes the total wave intensity and does not impact wave polarization (Wang et al. 2010). The wave mode coupling and polarization transfer in relativistic plasma have been studied by some authors (Cheng & Ruderman 1979; Wang et al. 2010). As the wave propagates in the magnetosphere, plasma density decreases, gradually diminishing its impact on the polarization of the waves. Wave mode coupling occurs near the "polarization limiting radius," $r_{pl}$, where the polarization is fixed, and the mode evolution changes from adiabatic to nonadiabatic. Wave mode coupling is generally induced by the rotation of the neutron star. The sign of CP will not change, and the polarization angle varies slowly (Wang et al. 2010).

The QT effect plays a crucial role in magnetospheric propagation. As the wave propagates through the magnetosphere, it traverses the QT region, where its wave vector becomes aligned or nearly aligned with the magnetic field (Wang & Lai 2009; Wang et al. 2010). In such a QT region, the impact angle $\chi$ is very small, causing the angle $\theta_B$ (between the wave vector and the magnetic field) to reach a minimum value at a large radius. A rapid change in the magnetic field's azimuthal angle results in the two photon modes (X-mode and O-mode) becoming nearly identical. The final polarization state after traversing this QT region is complicated: its polarization modes may experience temporary recoupling, resulting in partial mode conversion. PAs can undergo significant modifications, and distinct signs of CP may be generated for different geometries.

If the neutron star is nonrotating, the angle $\theta_B$ will increase monotonically, and there will be no QT propagation effect. Nonetheless, if the neutron star's rotation is considered, the angle $\theta_B$ would get its minimum value for a small impact angle $\chi$. After traveling through the QT region, the PAs vary significantly, and different signs of CP can be produced for different geometries (Wang et al. 2010). The polarization properties are different from the cyclotron absorption and wave mode coupling effect. FRB emission traveling through the open field line region of the magnetosphere of a magnetar does not suffer significant loss when $\theta_B$ is extremely small (Qu et al. 2022). In scenarios where $\theta_B \ll 1$, FRBs remain transparent within a magnetar's magnetosphere, even for high-luminosity FRBs characterized by large pair multiplicity, as discussed in Qu et al. (2023). Therefore, the QT propagation effect is beneficial to the propagation of FRBs. Nevertheless, when an FRB attempts to penetrate the closed field line region, the value of $\theta_B$ increases, which significantly enhances the scattering optical depth. Interestingly, a radio pulsar phase was detected in SGR J1935+2154, with the radio pulses originating from a fixed region within the magnetosphere, but bursts occurring in random locations (Zhu et al. 2023). If the impact angle is sufficiently large, FRBs can also be generated from the closed field lines region of a magnetar's magnetosphere. A geometric model for repeaters and apparent nonrepeaters is described in detail in the Appendix.

Assuming that both repeaters and apparent nonrepeaters originate from the same emission mechanism, the QT propagation effect within the magnetosphere can be advantageous in distinguishing their characteristics. In this model, the apparent nonrepeaters arise from sources with an emitting region closer to the magnetic axis than repeaters (see Figure 1). The QT effect has a pronounced impact on polarization phase profiles, especially when the impact angle is extremely small. Utilizing polarization phase profiles can serve as an effective method for distinguishing between repeaters and apparent nonrepeaters. Besides, the repetition is a crucial physical quantity for determining the activity of bursts. The intrinsic repetition rate is linked to the triggering mechanism of FRBs, but the observed repetition rate is related to observational effects. Apparent nonrepeaters are situated close to the magnetic axis and occupy only a small fraction of the neutron star's surface, resulting in a low observable repetition rate (Section 3.9). Furthermore, within the framework of coherent curvature radiation (Kumar et al. 2017; Yang & Zhang 2018), we introduce the impact angle to characterize the key features of repeaters and apparent nonrepeaters, including bandwidth, duration, luminosity, and time–frequency downward-drifting profile. In Section 3, we illustrate how the QT effect can be utilized to differentiate between repeaters and apparent nonrepeaters based on observations.

We have discussed how a variety of crucial physical effects are associated with wave propagation through the magnetosphere. However, in many cases, these different effects are coupled and challenging to separate. Therefore, to generate the observed polarization profiles, numerical ray integrations are necessary to calculate the final wave polarization states. In our calculations, the impact angle is a critical parameter affecting the evolution of polarization profiles. For a specific neutron star, observations at different lines of sight (i.e., different impact angles $\chi$) would obviously yield distinct intensity and polarization profiles. The polarization behaviors are influenced by the QT effect persisting across varying neutron star and plasma parameters (e.g., $B_s$, $P$, $\eta$, $\gamma$). Changes in parameters only alter the position of the 90° jump in PA and the initial rotation phase, while the fundamental morphology of the emission beam remains unchanged.

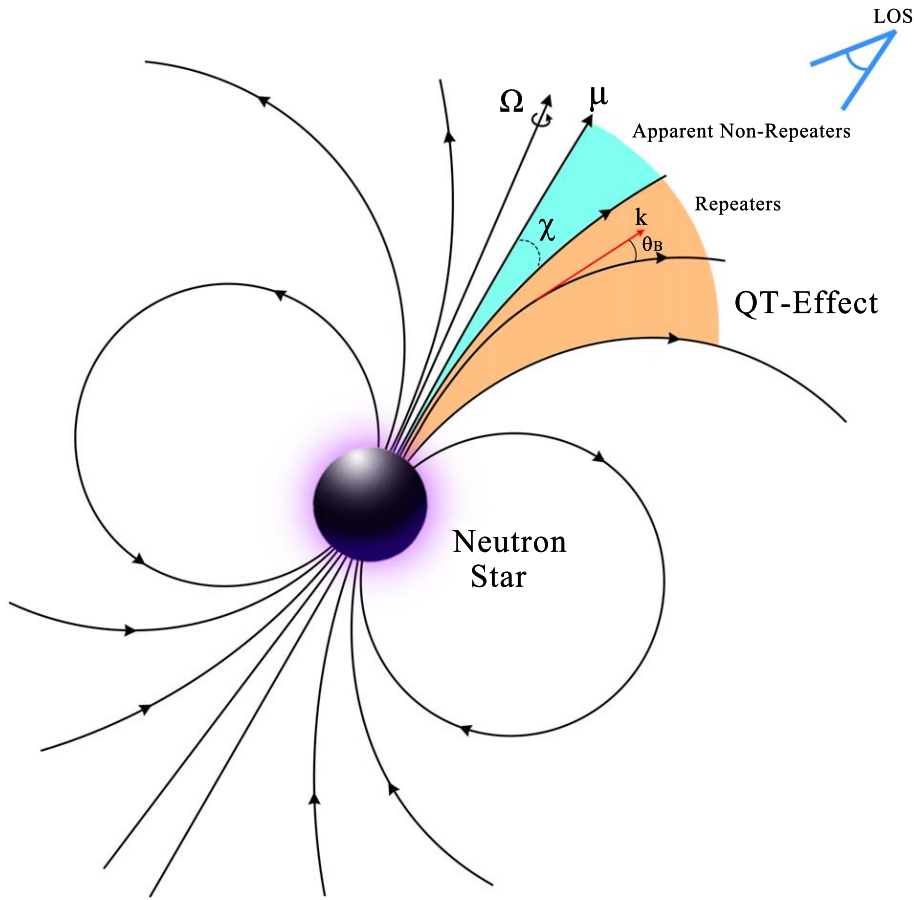

**Figure 1.** Geometric sketch of the model described in this paper. The QT propagation effect within the magnetosphere of the neutron star distinguishes repeaters and apparent nonrepeaters. The green region denotes apparent nonrepeaters with a smaller impact angle, and the orange region denotes repeaters with a larger impact angle. The angle between the LOS and the magnetic axis $\mu$ is represented by $\chi$, $\Omega$ is the rotation axis, and $\theta_B$ denotes the angle between the wave vector and magnetic field line.

## 3. Explanations for Observations of Repeaters and Apparent Nonrepeaters

From statistical analysis (e.g., Pleunis et al. 2021) or machine learning (e.g., B. H. Chen et al. 2022) using the data of CHIME or other radio telescopes, authors found that the observed bandwidth (Pleunis et al. 2021), duration, spectral index (Zhong et al. 2022), luminosity (Luo et al. 2020; H.-Y. Chen et al. 2022), energy (Hashimoto et al. 2020a, 2020b; Kim et al. 2022; Ikebe et al. 2023), and brightness temperature (Xiao & Dai 2022; Luo et al. 2023; Zhu-Ge et al. 2023) are other discriminant properties between repeaters and apparent nonrepeaters. Zhong et al. (2022) displayed a comprehensive statistical analysis using the first CHIME/FRB catalog. It identifies several distinctive properties distinguishing apparently nonrepeating FRBs from repeaters. On average, apparently nonrepeating FRBs exhibit a broader bandwidth ($308 \pm 103$ MHz) and higher energy ($\log(E/\text{erg}) = 40.0 \pm 1.1$) compared to repeaters, which have a bandwidth of $177 \pm 62$ MHz and energy $\log(E/\text{erg}) = 39.5 \pm 1.4$. They are also more luminous, with a log luminosity of $42.6 \pm 1.1$, in contrast to $41.7 \pm 1.3$ for repeaters. Despite a lower spectral index ($8.5 \pm 15.0$) compared to repeaters ($45 \pm 22$), apparently nonrepeating FRBs show a comparable peak frequency around $503 \pm 114$ MHz, slightly lower than that of repeaters $528 \pm 73$ MHz. The differences in polarization between repeating and apparently nonrepeating bursts are shown in Sections 3.1–3.3.

### 3.1. Circular Polarization

Observationally, the variation timescale of CP for apparent nonrepeaters is much smaller than that for repeaters (Cho et al. 2020; Day et al. 2020; Feng et al. 2022b). The degree of CP remains relatively constant in ∼2 ms timescale for FRB 20121102A, with no sign of change (Feng et al. 2022b). The observed significant CP is unlikely induced by multipath propagation (Feng et al. 2022b; Yang et al. 2022). In contrast, apparently nonrepeating FRBs exhibit a much shorter timescale for drastic variations in CP, typically less than 1 ms (Cho et al. 2020; Day et al. 2020). Surprisingly, the degree of CP of FRB 20181112A varies from −34% to 17% in less than 0.1 ms (Cho et al. 2020). The question of whether the polarization differences between repeaters and apparent nonrepeaters are a result of physically distinct origin is still unknown. Very recently, a negative correlation between the rotation measure and the CP fraction was discovered by FAST (Zhang et al. 2023), suggesting a potential connection between the origin of CP and the physics within the magnetosphere. The variation of the CP fraction could be caused by an intrinsic radiation mechanism (Wang et al. 2022a) or propagation effects (Beniamini et al. 2022). The QT propagation effect provides a crucial clue for understanding the evolution of CP.

We use numerical calculations to get the final wave polarization profiles. As shown in Figure 2, the evolution of the CP with phase for apparent nonrepeaters (a smaller $\chi$) is more drastic than that of repeaters (a larger $\chi$). We also present the fitting results of the CP evolution with phase (see Figure 3), which is consistent with the observations (Cho et al. 2020; Day et al. 2020; Feng et al. 2022b). The variation timescale of the CP for apparent nonrepeaters is much smaller than that of repeaters. As shown in the right-hand panel of Figure 2, a stronger magnetic field will cause a flatter evolution of the CP for a larger impact angle. Besides, one can see that the CP retains a single sign, which is the same as the case with a large impact parameter. However, the sign of CP will change for a very small impact angle since the QT propagation effect is much stronger. According to the evolution of polarization profiles, the impact angle of apparent nonrepeaters can be approximately constrained within 1° due to the drastic CP evolution. More data of the polarization evolution for repeaters and apparent nonrepeaters will provide more precise angle constraints. Therefore, the CP profile can be used as an effective way to distinguish repeaters and apparent nonrepeaters. Within the framework of coherent curvature radiation, if the bunches of particles are nonuniformly distributed, or the bunched opening angle $\sim 1/\gamma$, the circularly polarized bursts would be generated within the emission cone (Wang et al. 2022a; Liu et al. 2023b).

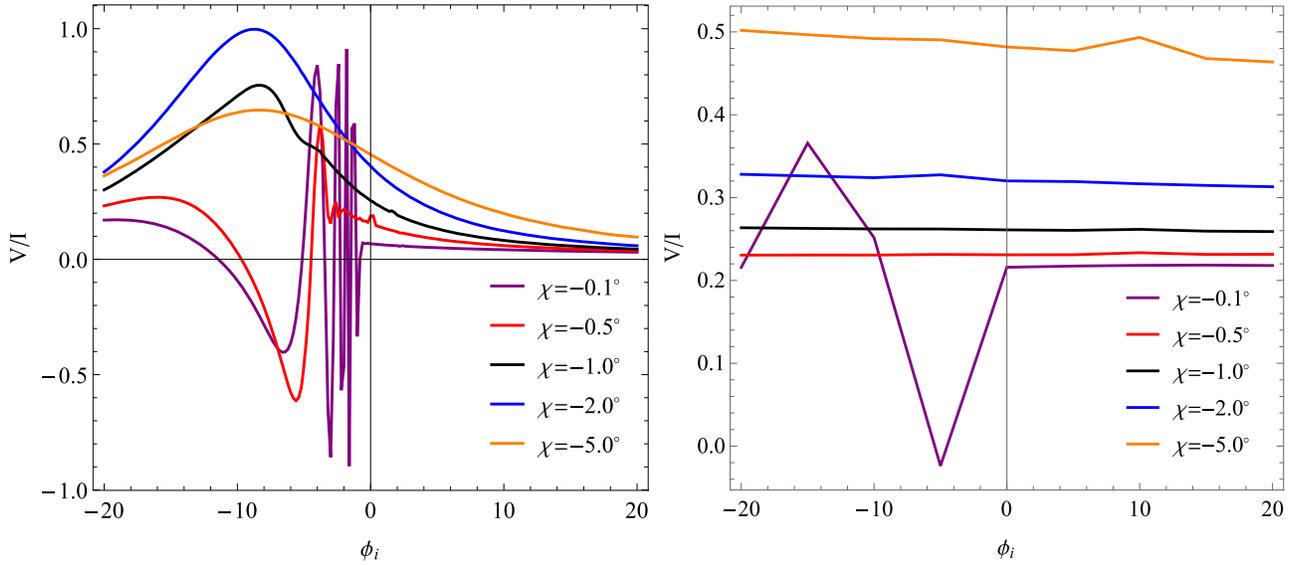

**Figure 2.** The evolution of CP with phase $\phi_i$. The purple line, red line, black line, blue line, and orange line indicate $\chi = -0^\circ\!.1$, $-0^\circ\!.5$, $-1^\circ\!.0$, $-2^\circ\!.0$, and $-5^\circ\!.0$, respectively. The left-hand panel denotes $B_s = 10^{12}$ G and the right-hand panel denotes $B_s = 10^{15}$ G. It should be noted that the sign of $\chi$ indeed influences the sign of CP. We consider the following parameters: the positive and negative electrons move with the same single Lorentz factor $\gamma = 100$, the multiplicity $\eta = 400$, and the period of the neutron star $P = 1$ s.

The polarization observations of radio emission from magnetars could provide a vital clue for researching the origins of FRBs. XTE J1810-197 was the first transient and radio-emitting magnetar (Camilo et al. 2006). After spending almost a decade in a quiescent, radio-silent state, the magnetar was reported to have undergone a radio outburst in 2018 December (Lyne et al. 2018). A high time-resolution polarimetric single pulse profile with millisecond duration was observed by Dai et al. (2019). Spikes within the latter part of the main pulse and tail components are almost completely linearly polarized, but most high circularly polarized bursts are generated within the first half of the main pulse. Interestingly, the fine structure is very similar to our result (see Figure 2). It is suggested that the QT propagation effect in the magnetosphere of a magnetar plays an important role in the evolution of CP. If the fine structure of polarization for repeaters and apparent nonrepeaters can be detected by FAST or other telescopes in the future, this will further provide a smoking gun for the origins of repeaters and apparent nonrepeaters.

### 3.2. Polarization Angles

The PAs keep nearly constant across each burst for most repeating FRBs (Michilli et al. 2018; CHIME/FRB Collaboration et al. 2019). However, the repeater FRB 180301 shows PAs that vary remarkably with time (Luo et al. 2020). Besides, the majority of bursts for the repeater FRB 20201124A show a nonvarying PA, and a significant PA variation is detected in a small portion of bursts (Jiang et al. 2022). For some apparent nonrepeaters, a swing of PAs across each burst is observed, and the swing profiles are extraordinarily diverse among bursts (Masui et al. 2015; Ravi et al. 2016; Prochaska et al. 2019; Cho et al. 2020; Day et al. 2020). An S- or inverse S-shaped pattern can usually be observed in radio pulsars (Lorimer & Kramer 2012), and the patterns have been predicted by the rotating vector model (Radhakrishnan & Cooke 1969). The evolution of PAs is sensitive to the QT propagation effect in our model. As shown in Figure 4, apparent nonrepeaters exhibit more drastic evolution in PAs over a shorter timescale compared to repeaters. We find that the PA profile is similar to the prediction from the rotating vector model for a relatively large impact parameter, which corresponds to a flat PA evolution within the burst phases. Nonetheless, the PA would show a sudden 90° jump and a diverse swing profile for a smaller impact parameter due to the QT propagation effect. Most apparent nonrepeaters display a remarkable PA variation (Masui et al. 2015; Ravi et al. 2016; Prochaska et al. 2019; Cho et al. 2020; Day et al. 2020), which might hint that the QT propagation effect is very strong. Furthermore, the variation in PA for both repeaters and apparent nonrepeaters might be attributed to the intrinsic radiation mechanism. Within the framework of coherent curvature radiation, the majority of bursts have an on-beam case (i.e., the LOS is within the bunched opening angle) to produce the small PA variation (Liu et al. 2023b).

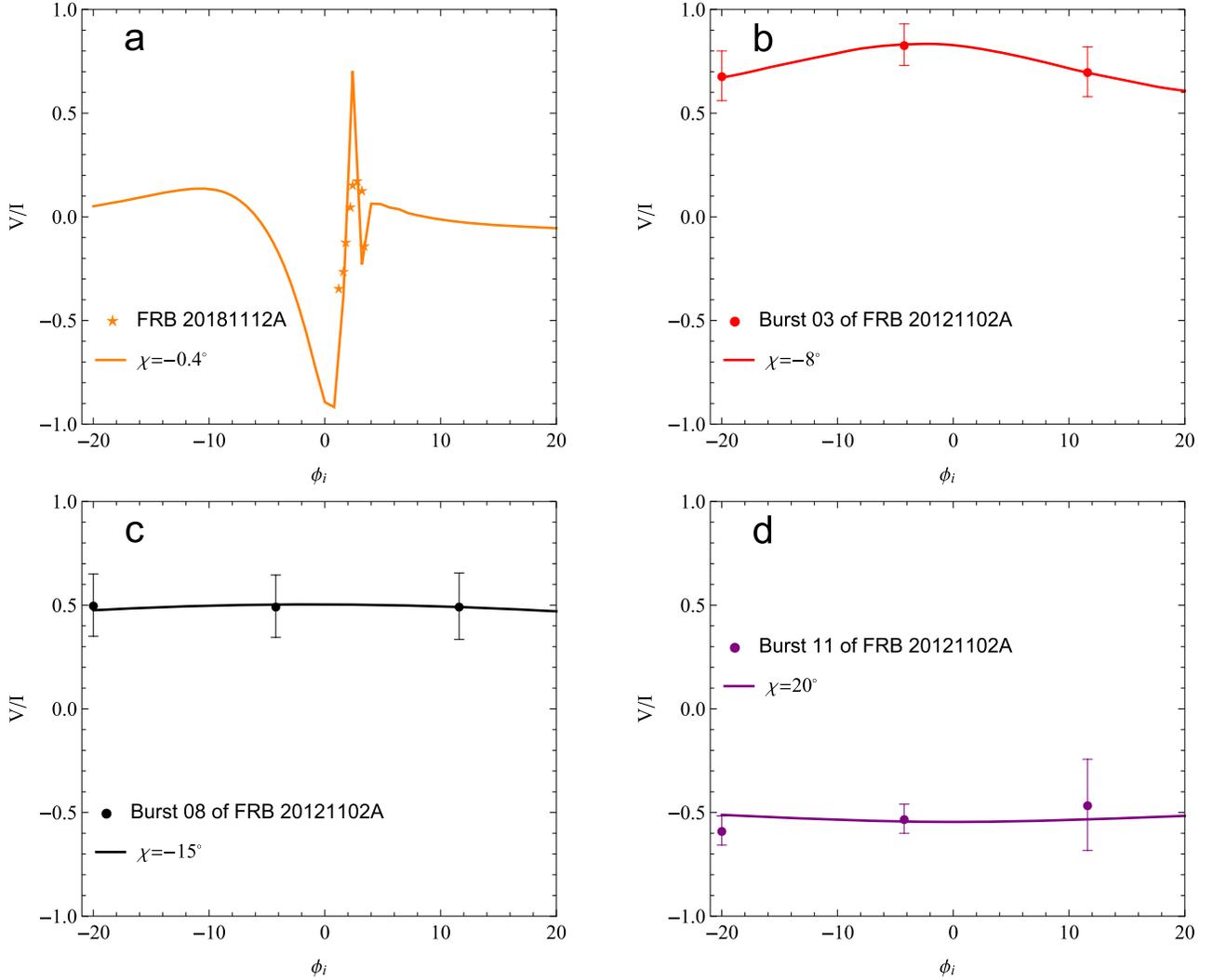

**Figure 3.** The fitting results of the CP evolution with phase. Panel (a): the orange stars show the data of apparent nonrepeater FRB 20181112A (Cho et al. 2020), and the orange line denotes the fitting result with $\chi = -0°.4$. Panels (b), (c), and (d): the red points, black points, and purple points denote the burst 03, 08, and 11 of repeater FRB 20121102A, respectively (Feng et al. 2022b). The red line, black line, and purple line show the fitting results with $\chi = -8°$, $\chi = -15°$, and $\chi = 20°$, respectively. We consider the following parameters: the positive and negative electrons move with the same single Lorentz factor $\gamma = 100$, the multiplicity $\eta = 400$, the period of the neutron star $P = 1$ s, and $B_s = 10^{12}$ G.

### 3.3. Linear Polarization

Linear polarization (LP) fractions are near 100% across each pulse for most repeaters (CHIME/FRB Collaboration et al. 2019; Nimmo et al. 2021). However, a few repeaters (e.g., FRB 180301) show partial LP (Cho et al. 2020; Luo et al. 2020). Most of the apparent nonrepeaters have relatively high LP fractions (Masui et al. 2015; Ravi et al. 2016). Very recently, a significant trend of increased depolarization at lower frequencies was observed for five active repeaters (Feng et al. 2022a). This can be well explained by the multipath propagation through a magnetized inhomogeneous plasma screen (Yang et al. 2022). Within the framework of curvature radiation, most of the depolarization degrees of bursts have a small variation in a wide frequency band due to the slow evolution of LP as a frequency (Liu et al. 2023b). As shown in Figure 5, one can see a clear trend of higher LP at higher frequencies, and the impact angle affects the intensity of LP for the QT propagation effect within the magnetosphere. The frequency-dependent LP degree is consistent with the observation of Feng et al. (2022a). Thus, the QT propagation effect within the magnetosphere could also contribute to depolarization.

### 3.4. Bandwidth

As mentioned before, apparent nonrepeaters are from bursts occurring in an emitting region near the magnetic axis. In this case, the emitting region relative to apparent nonrepeaters should be threaded by magnetic field lines with a larger curvature radius than that of repeaters due to the relation between the curvature radius $\rho$ and the impact angle $\chi$ with respect to the magnetic axis (e.g., Kumar et al. 2017):

$$\rho(R, \chi) \approx \frac{0.8R}{|\chi|}, \quad (1)$$

where $R$ is the radius with respect to the stellar center. This is the key intrinsic distinction giving rise to the observed apparent nonrepeaters and repeaters in this geometric model. To analyze how this key distinction leads to the observed discriminant properties, one should consider the specific observational selection effects of telescopes and the coordinate transformation between the lab frame and the observed frame. For example, one burst can be observed by the CHIME only if its observer-frame frequency lies in 400–800 MHz. Moreover, for the observed apparent nonrepeaters and repeaters from the CHIME, no notably different peak frequency (i.e., characteristic frequency) distributions were found (Zhong et al. 2022). If this is the case, the apparent nonrepeaters and repeaters can be assumed to share a comparable observer-frame characteristic frequency of curvature radiation $\nu_{obs}$ lying in 400–800 MHz, which is linked to that of the lab frame $\nu_{lab} = \frac{3c\gamma^3}{4\pi\rho}$ (Jackson et al. 1998) via (e.g., Equation (3.46) in Zhang 2018)

$$\nu_{obs} \simeq 2\gamma^2 \nu_{lab} = \frac{3c\gamma^5}{2\pi\rho} = \frac{3c\gamma^5|\chi|}{1.6\pi R}, \quad (2)$$

where $\gamma$ is the common Lorentz factor of charges. Therefore, the apparent nonrepeaters with a smaller impact angle should possess a larger intrinsic common Lorentz factor if the emitting region radius $R$ is substantially comparable for all FRBs. Similarly, the observed bandwidth can be given by

$$\Delta\nu_{obs} \simeq \frac{3c\gamma^5}{1.6\pi R}|\Delta\chi| = \nu_{obs}\left|\frac{\Delta\chi}{\chi}\right|. \quad (3)$$

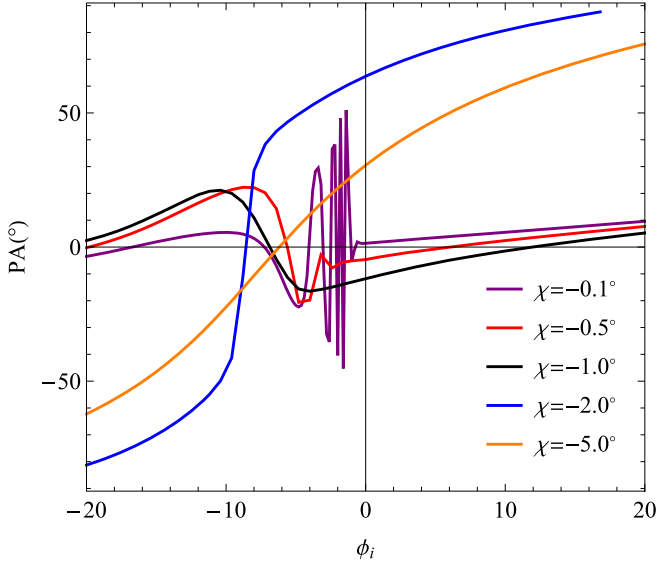

**Figure 4.** The evolution of PA with phase $\phi_i$. The purple line, red line, black line, blue line, and orange line indicate $\chi = -0°.1, -0°.5, -1°.0, -2°.0$, and $-5°.0$, respectively. We consider the following parameters: the positive and negative electrons move with the same single Lorentz factor $\gamma = 100$, the multiplicity $\eta = 400$, the period of the neutron star $P = 1$ s, and $B_s = 10^{12}$ G.

If the change in impact angle, $\Delta\chi$, is comparable between apparent nonrepeaters and repeaters, then apparent nonrepeaters, having a smaller impact angle, would exhibit an observed broader bandwidth compared to repeaters. This is consistent with observations reported by Pleunis et al. (2021).

### 3.5. Duration

From the above results, apparent nonrepeaters naturally have an observed shorter duration than repeaters because the observer-frame duration $\tau_{obs}$ is related to the lab frame $\tau_{lab}$

through (e.g., Equation (3.35) in Zhang 2018)

$$\tau_{obs} = \frac{\tau_{lab}}{2\gamma^2}, \quad (4)$$

if the lab-frame duration is roughly the same for all FRBs. Alternatively, the larger Lorentz factor of the apparent nonrepeaters, in comparison with repeaters, should be related to a narrower beam solid angle $\Omega$, within which electromagnetic waves add up coherently because of $\Omega \sim \pi\eta^{-1}\gamma^{-2}$ (Kumar et al. 2017) where $\eta = \eta_x\eta_y$ contains two multiplication factors $\eta_x$ and $\eta_y$ corresponding to the directions of the two principal axes in the transverse plane. If the FRB duration is connected with the timescale of the beam emission sweeping past the observer, apparent nonrepeaters with a narrower beam would sweep past the observer in a shorter time than repeaters, given that the angular speed at which the beam drifting past the observer is substantially the same for all FRBs (Connor et al. 2020). In this scenario, the apparent nonrepeaters would have a shorter observed duration distribution.

### 3.6. Luminosity and Energy

Based on the observer-frame isotropic luminosity calculation for coherent curvature radiation (e.g., Equation (20) in Kumar et al. 2017), one obtains

$$\begin{aligned} L_{iso,obs} &\sim 10^{41}\ \mathrm{erg\ s^{-1}}\ N_{patch}\eta^2\nu_{lab,9}^{-4/3}\rho_5^{8/3}(n'_{e,15})^2 \\ &\sim 3\times 10^{38}\ \mathrm{erg\ s^{-1}}\ N_{patch}\eta^2\xi^2 P_0^{-2} B_{s,15}^2 R_{s,6}^6 \\ &\quad \times \nu_{obs,9}^{-4/5} R_7^{-14/5} |\chi|^{-16/5}, \end{aligned} \quad (5)$$

where $N_{patch}$ is the number of coherent patches and $n'_e = \xi n_{GJ}$, in which $n_{GJ} \sim 10^{14}\ \mathrm{cm^{-3}}\ P_0^{-1} B_{s,15}(R_s/R)^3$ is the Goldreich–Julian density (Goldreich & Julian 1969); here, $P$, $B_s$, and $R_s$ are the spin period, surface magnetic field, and radius of the neutron star, respectively. As shown in Equation (5), the impact angle has a significant impact on luminosity, making it a sensitive discriminator between repeaters and apparent nonrepeaters based on their luminosities. In arriving at the above equation, we have used Equations (15) and (16) in Kumar et al. (2017), as well as Equations (1) and (2). Similarly, the observer-frame isotropic energy can be estimated by using Equations (4) and (5); one then has

$$\begin{aligned} E_{iso,obs} &\simeq L_{iso,obs}\tau_{obs} \\ &\sim 1.5\times 10^{35}\ \mathrm{erg}\ N_{patch}\eta^2\xi^2 P_0^{-2} B_{s,15}^2 R_{s,6}^6 \\ &\quad \times \nu_{obs,9}^{-6/5} R_7^{-16/5} \tau_{lab,-3} |\chi|^{-14/5}. \end{aligned} \quad (6)$$

As we can see, as long as apparent nonrepeaters and repeaters have an observed comparable characteristic frequency as well as the same radiation radius and intrinsic lab-frame duration, the former, typically with smaller impact angles, naturally have observed higher luminosity and energy than the latter, on average.

### 3.7. Brightness Temperature

Observed brightness temperature relates to the observed flux density $F_\nu$, characteristic frequency $\nu_{obs}$, and duration $\tau_{obs}$ (e.g.,

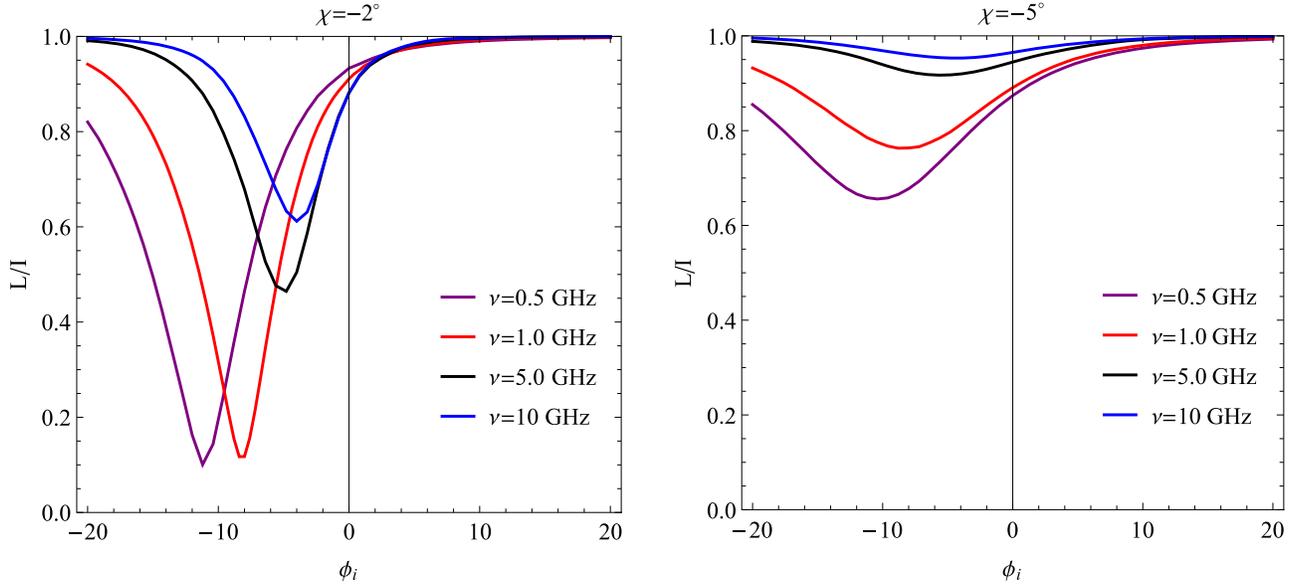

**Figure 5.** The evolution of LP with frequency. The purple line, red line, black line, and blue line indicate $\nu = 0.5$ GHz, 1.0 GHz, 5.0 GHz, and 10 GHz, respectively. The left-hand panel denotes $\chi = -2°$ and the right-hand panel denotes $\chi = -5°$. We consider the following parameters: the positive and negative electrons move with the same single Lorentz factor $\gamma = 100$, the multiplicity $\eta = 400$, the period of the neutron star $P = 1$ s, and $B_s = 10^{12}$ G.

Luo et al. 2023) through

$$T_{b,obs} = \frac{F_{\nu,obs} D_L^2}{2\pi k_B \nu_{obs}^2 \tau_{obs}^2 (1+z)}$$
$$\sim 10^{32} \text{ K } (1+z)^{-1} N_{patch} \eta^2 \xi^2 P_0^{-2} B_{s,15}^2 R_{s,6}^6$$
$$\times \tau_{lab,-3}^{-2} \nu_{obs,9}^{-3} R_7^{-2} |\chi|^{-3} |\Delta\chi|^{-1}, \quad (7)$$

where $F_{\nu,obs} = \frac{L_{iso,obs}}{4\pi D_L^2 \Delta\nu_{obs}}$; here, we have used Equations (3)–(5). When $\nu_{obs}$, $R$, $\tau_{lab}$, and $|\Delta\chi|$ are all comparable between the apparent nonrepeaters and repeaters, one can see that apparent nonrepeaters with a smaller impact angle would have a higher brightness than repeaters, consistent with the statistical results.

### 3.8. Time–Frequency Downward Drifting

The observed time–frequency from most repeaters is downward drifting (Gajjar et al. 2018; Michilli et al. 2018; Hessels et al. 2019; Josephy et al. 2019; Caleb et al. 2020; Day et al. 2020; Platts et al. 2021; Pleunis et al. 2021). This implies that it is very likely a common property for repeaters, whereas most apparent nonrepeaters have almost no frequency drift profile, or a few bursts show the upward/downward-drifting pattern (Pleunis et al. 2021). Explanations for the drifting profile have been done either in magnetosphere models or blast wave models (Metzger et al. 2019; Wang et al. 2019). If the downward drifting is created from the magnetosphere, it is a natural consequence of bunched coherent curvature radiation due to the radius-to-frequency mapping (Wang et al. 2019, 2022b). If the trigger event has a longer duration, there would be a higher probability to observe upward-drifting events (Wang et al. 2020). The drifting rate varies with frequency, which can be described by Wang et al. (2020):

$$\dot{\nu} = \nu \left[ 3\frac{\Delta\gamma}{\gamma \Delta t_{obs}} - \frac{\Delta\rho}{\rho \Delta t_{obs}} \right], \quad (8)$$

where $\Delta t_{obs}$ is the observed time delay of the two subpulses. According to Equation (2), we can get the relation of $\gamma = (2\pi\rho\nu_{obs}/3c)^{1/5}$. One obtains $\dot{\nu} \propto \Delta\rho^{1/5}\rho - \rho^{1/5}\Delta\rho$. Combining Equation (1), the drifting rate can be given by $\dot{\nu} \propto |\Delta\chi|^{-1/5}|\chi|^{-1} - |\chi|^{-1/5}|\Delta\chi|^{-1}$. When $\nu_{obs}$, $R$, and $\Delta t_{obs}$ are all comparable between the apparent nonrepeaters and repeaters, one can see that $|\Delta\chi| \simeq |\chi|$ for most apparent nonrepeaters due to a smaller $\chi$, and $|\Delta\chi| \lesssim |\chi|$ for most repeaters. Thus, most apparent nonrepeaters have almost no frequency drift profile. Even in cases where $|\Delta\chi|$ is larger than $|\chi|$, an upward-drifting profile can be observed. Conversely, most repeaters display a downward-drifting pattern. Besides, the extremely short duration of apparent nonrepeaters could be why the drifting profile is difficult to observe.

### 3.9. Repetition Rates

Why are there so many apparent nonrepeaters (more than 800) but very little repetition? The selection effect of observations may play an important role. Some apparent nonrepeaters may have actually repeated, but the telescope missed those repetitions, or they repeated at a level below the telescope's sensitivity, and so on. The intrinsic repetition rate is linked to the triggering mechanism of FRBs, but the observed repetition rate is related to observational effects. If FRBs originate from the active regions of a neutron star, this could lead to a higher intrinsic repetition rate. Since it is challenging to pinpoint these active regions, it is reasonable to assume that the active regions are randomly distributed on the surface of a neutron star. The observable probability of each surface element on the surface of a neutron star is the same. Apparent nonrepeaters are situated close to the magnetic axis and occupy only a small fraction of the neutron star's surface, resulting in a low observable repetition rate. Furthermore, a lower repetition rate suggests a narrower beaming angle, which can be attributed to the larger Lorentz factor of the apparent nonrepeaters. Based on the assumption of Poissonian repetition by Connor et al. (2020), we consider that, on average, the probability of detecting an apparent nonrepeater from a given source is $|\chi|/\pi$. Assuming Poissonian repetition where the $j$th source has a repetition rate $\mathcal{R}_j$, opening angle $\Omega_j$, and impact angle $\chi_j$, the expected number of bursts in observing time $T_{\rm obs}$ is

$$N_{\rm exp}^j = \frac{\Omega_j}{4\pi}\frac{|\chi_j|}{\pi}\mathcal{R}_j T_{\rm obs}, \quad (9)$$

the probability of a source being detected once can be written as

$$P(n=1|\Omega_j, \chi_j, \mathcal{R}_j) = e^{-N_{\rm exp}^j} N_{\rm exp}^j, \quad (10)$$

and the probability of a source being detected more than once can be given by

$$P(n \geqslant 2|\Omega_j, \chi_j, \mathcal{R}_j) = 1 - e^{-N_{\rm exp}^j} - P(n=1|\Omega_j, \chi_j, \mathcal{R}_j). \quad (11)$$

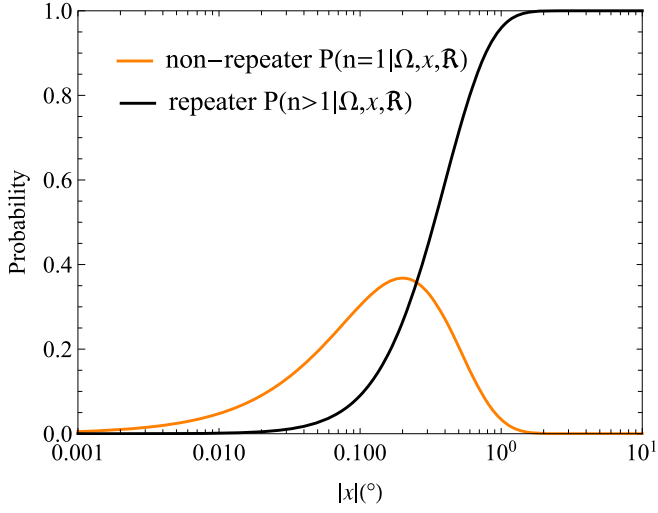

**Figure 6.** The probability of a beamed FRB emitting in the direction of an observer once (orange) and more than once (black) as a function of the impact angle $|\chi|$.

As shown in Figure 6, the probability of a beamed FRB emitting once in the direction of an observer has a smaller impact angle distribution than those emitting more than once. Note that the impact angle $|\chi|$ of apparent nonrepeaters can be approximately constrained within 1° according to the evolution of polarization profiles. Consider that the polar cap of the pulsar is enclosed within the last open field lines with a polar angle $\theta_p \simeq 0.01 P_0^{-1/2}$. The solid angle of polar cap $\Omega_p \sim 0.5 \theta_p^2$, and the solid angle of apparent nonrepeater can be constrained within $\Omega_{\rm non-repeater} \sim 0.5(\pi/180)^2$. Thus, the solid angle of the repeater can be given by $\Omega_{\rm repeater} = \Omega_p - \Omega_{\rm non-repeater}$. As shown in Figure 7, we simulate the solid angle distribution for the apparent nonrepeaters and repeaters. One can see that apparent nonrepeaters have a smaller solid angle distribution compared to repeaters. This is a likely reason why apparent nonrepeaters exhibit a lower observed repetition rate than repeaters.

Many apparent nonrepeaters may repeat later as the impact angle increases. Interestingly, some bursts of repeaters show some characteristics of apparent nonrepeaters. The observation features (duration, energy density) of first-detected repeaters are consistent with apparent nonrepeaters (Ikebe et al. 2023). Most of the bursts from the repeater FRB 20201124A show a nonvarying PA and low CP, while a small fraction of bursts display diverse PA variations and high CP (Jiang et al. 2022). Besides, repeater FRB 180301 shows that PAs vary remarkably with time (Luo et al. 2020). Given the randomness of the LOS, a large impact angle becomes highly probable, especially when repeaters can be accurately located and monitored over an extended period. Furthermore, changes in the rotation axis can also contribute to variations in the impact angle. Consequently, some bursts from repeaters may exhibit characteristics typical of apparent nonrepeaters. Our model provides a coherent interpretation of the observations of both repeaters and apparent nonrepeaters.

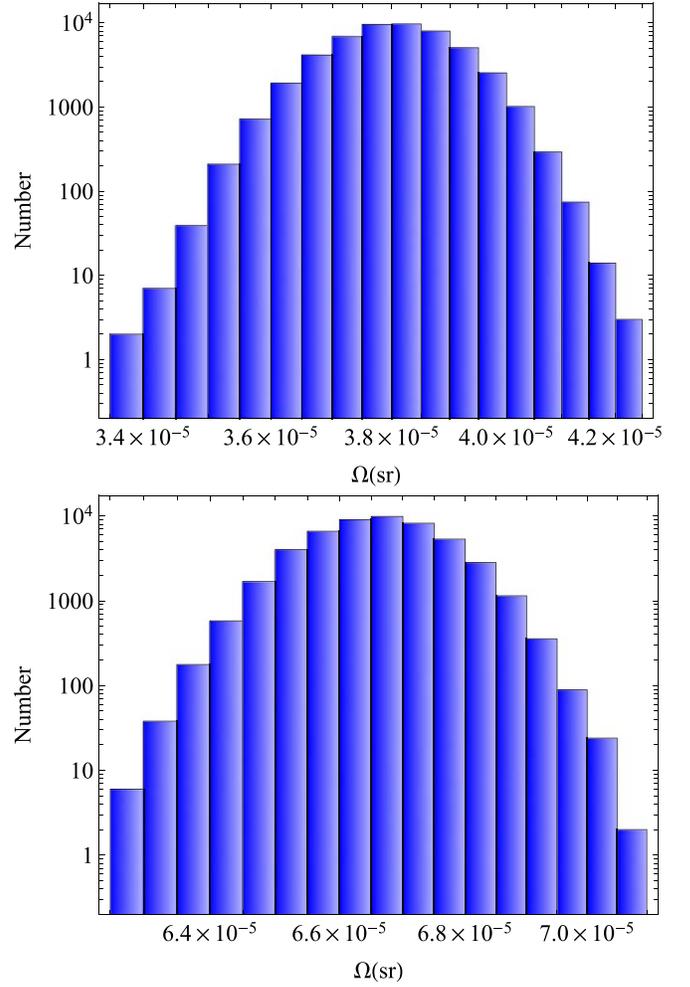

**Figure 7.** Simulated solid angle distribution for the apparent nonrepeaters (upper panel) and repeaters (bottom panel).

## 4. Conclusions and Discussion

In this paper, we proposed a unified geometric model to distinguish between repeaters and apparent nonrepeaters, in which the QT propagation effect within the magnetosphere of a neutron star is considered. In this model, apparent nonrepeaters arise from sources whose emitting region has a smaller impact angle with respect to the magnetic axis, while repeaters come from sources whose emitting region has a larger impact angle. Notably, nearly all distinguishing characteristics of repeaters and apparent nonrepeaters can be effectively understood in our model. Apparent nonrepeaters display more drastic CP and PA evolution in a shorter timescale, broader bandwidth, shorter duration, higher peak luminosity, greater energy, elevated brightness temperature, reduced drifting rate, and lower observed repetition rate compared to repeaters.

Observationally, many repeaters might be observed as apparent nonrepeaters. Repeaters can be as bright as apparent nonrepeaters, and only their brightest bursts might be observed, and thus some repeaters intrinsically might have been classified as apparent nonrepeaters (Ikebe et al. 2023). CHIME has only once detected the repeater FRB 20121102A; if there were no previous observations, this source would be classified as an apparent nonrepeater (Josephy et al. 2019). Thus, some one-off events are part of repeaters. Interestingly, we find that apparent nonrepeaters and repeaters could correspond to the core region and hollow cone region in the polar region of the pulsar, respectively (Backer 1976). For an extremely small impact angle, aligning with the core emission of pulsars, the QT effect can lead to a sign reversal of CP, as observed in the core components of some pulsars (e.g., Radhakrishnan & Rankin 1990). Monitoring other repeaters and deep follow-up observations of apparent nonrepeaters are needed to understand the origins of repeaters and apparent nonrepeaters.

Based on observations, the number of detected apparently nonrepeating bursts is greater than that of repeating bursts (Xu et al. 2023). The event rate of FRBs is influenced by a variety of physical factors, including the activity of the sources, among others. Although apparently nonrepeating bursts occur at small impact angles, we assume that such sources are numerous, meaning that the majority of sources experience bursts at small impact angles. Additionally, the lower event rate of repeating bursts may be related to the effects of propagation in the magnetosphere. FRB emission traveling through the open field line region of a magnetar's magnetosphere does not suffer significant loss when $\theta_B$ is extremely small (Qu et al. 2022). Radio waves propagating parallel to the magnetic field lines are more likely to escape the magnetosphere, making bursts near the magnetic axis more favorable for meeting the conditions of parallel propagation. Therefore, the number of apparently nonrepeating bursts is expected to be greater than that of repeating bursts.

The spectrum profile can be used to test the emission mechanism of FRBs. Many observations infer that the intrinsic spectrum of FRBs is narrow (Lorimer et al. 2007; Masui et al. 2015). Interestingly, apparent nonrepeaters tend to have a relatively small spectral index and a relatively flat spectrum, while repeaters have an extremely steep spectrum (Zhong et al. 2022). For coherent curvature radiation, the spectrum is found to be a broken power law separated by some characteristic frequencies (Yang & Zhang 2018). The narrow spectrum could be generated by the charge separation (Yang et al. 2020) or the coherent process by multiple particles (Yang 2023) within the framework of coherent curvature radiation. For coherent inverse Compton scattering, the characteristic frequency has a narrow value range (i.e., the oscillation frequency, the Lorentz factor, and the photon incident angle change very little), suggesting that it is an intrinsically narrowband mechanism (Zhang 2022). Very recently, bunched coherent Cherenkov radiation was proposed as the radiation mechanism of FRBs by Liu et al. (2023). Some relativistic particles are emitted from the polar cap of a magnetar and move along the field lines through a charge-separated magnetic plasma, emitting coherent Cherenkov radiation along the way. The bunched coherent Cherenkov radiation can generate an extremely narrowband spectrum. More observational data on repeaters and apparent nonrepeaters in the future will help test these radiation mechanisms.

Polarization of normal pulsars, millisecond pulsars, and magnetars with radio emission may provide a significant clue for studying the origins of FRBs. Most pulsars (normal and millisecond) show that the CP degree varies from 5% to 20% (Dai et al. 2015). However, the CP of magnetars is generally larger than normal and millisecond pulsars (Dai et al. 2019). The observations suggest that a complex magnetic field configuration or propagation effects within the magnetosphere could affect the CP (Dai et al. 2021; Tong et al. 2021). Besides, quick variations in PAs and the sign of CP have been detected. A very bright FRB-like burst from a Galactic magnetar SGR J1935+2154 was reported (Bochenek et al. 2020; CHIME/FRB Collaboration et al. 2020). A strong LP and no CP were observed in FRB 200428 (CHIME/FRB Collaboration et al. 2020). However, some radio bursts from SGR J1935+2154 were observed by Kirsten et al. (2021) after a few days. Those bursts show a significant CP and PA variations (Kirsten et al. 2021). This indicates that the propagation effect, with the QT effect being primary, within the magnetosphere of SGR J1935+2154 plays a vital role in the evolution of CP and PAs.

Some intrinsic factors could affect the properties of polarization. The significant difference in duration and repetition distributions between repeaters and apparent nonrepeaters can be attributed to the selection effect owing to a beamed emission (Connor et al. 2020). Repeaters have a larger opening angle to contribute to a longer pulse width. The large opening angle makes most repeaters have a high LP degree. We find this is consistent with the simulation results of repeaters by Liu et al. (2023b). Within the framework of coherent curvature radiation, the bunched opening angle will significantly influence the polarization of emission, and most bunches have large opening angles to contribute to the high LP distribution (Liu et al. 2023b). Additionally, the PA variation is correlated with the CP for curvature radiation. Most low CP bursts and small PA variation are generated within the emission cone (Liu et al. 2023b), while the PA variation is also correlated with the CP for propagation effects within the magnetosphere. Comparing the profile and statistical distribution of polarization can help us to distinguish the intrinsic radiation mechanism and propagation effects. Our future work will further distinguish between the inherent radiation mechanism and propagation effects.

## Acknowledgments

We are grateful to Chen Wang, Yuan-Pei Yang, Wei-Yang Wang, Jin-Jun Geng, Zi-Ke Liu, and an anonymous referee for helpful discussions and constructive suggestions. This work was supported by the National SKA Program of China (grant Nos. 2020SKA0120300 and 2022SKA0130100), the National Natural Science Foundation of China (grant Nos. 12393812, 12273009, 12273009, 12247144 and U2038105), the Program for Innovative Talents, Entrepreneur in Jiangsu, and China Postdoctoral Science Foundation (grant Nos. 2021TQ0325 and 2022M723060).

## Appendix

We consider a photon (radio wave) radiated at time $t_i$ and position $r_i$ and assume that the photon's trajectory is a straight line along the wave vector $k$, i.e., the direction of photon propagation remains unchanged because we ignore the refractive effect of the photon. We choose a coordinate system *XYZ* (see Figure 8), where the LOS is along the *Z*-axis, and the rotation axis $\mathbf{\Omega}$ is located in the *XZ* plane. At a given time, we can write the photon's position and the corresponding pulsar rotation phase:

$$\boldsymbol{r} = \boldsymbol{r}_i + s\hat{z}, \quad (A1)$$

$$\phi = \phi_i + \Omega(t - t_i) = \phi_i + s/r_{lc}, \quad (A2)$$

where $s = c(t - t_i)$ is the propagation distance of the photon and $r_{lc} = c/\Omega$ is the light cylinder radius. For a dipole field, in the coordinate system $X'Y'Z'$ with the magnetic moment $\boldsymbol{\mu}$ along $Z'$, the magnetic field can be given by

$$B'_x = \frac{3B_s R_s^3}{2r^3} \sin\theta \cos\theta \cos\varphi, \quad (A3)$$

$$B'_y = \frac{3B_s R_s^3}{2r^3} \sin\theta \cos\theta \sin\varphi, \quad (A4)$$

$$B'_z = \frac{B_s R_s^3}{2r^3}(2\cos^2\theta - \sin^2\theta). \quad (A5)$$

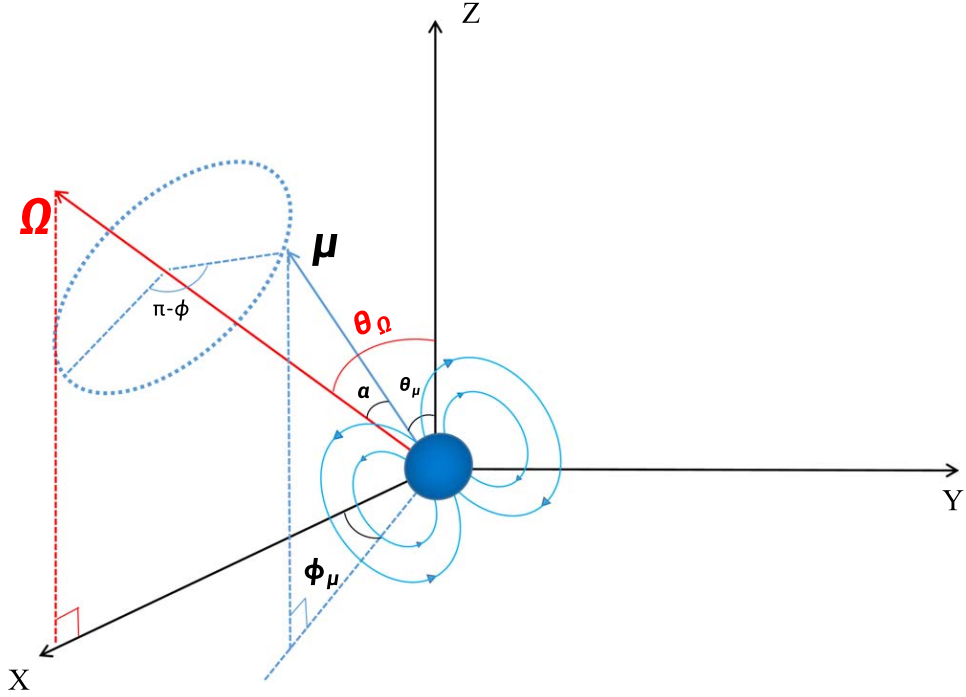

**Figure 8.** The frame used in this paper. We choose a coordinate system *XYZ* such that the LOS is along the *Z*-axis and the rotation axis $\mathbf{\Omega}$ is located in the *XZ* plane. $\alpha$ is the magnetic inclination between the magnetic axis $\boldsymbol{\mu}$ and the axis of rotation, $\theta_\Omega$ is the angle between the axis of rotation and the *Z*-axis, and the direction of the magnetic axis in the *XYZ* coordinate system can be expressed by the polar angle $\theta_\mu$ and azimuthal angle $\phi_\mu$.

The direction of the magnetic axis in the *XYZ* coordinate system can be expressed by the polar angle $\theta_\mu$ and azimuthal angle $\phi_\mu$, which can be given by

$$\tan\phi_\mu = \frac{\sin\alpha \sin\phi}{\cos\alpha \sin\theta_\Omega + \sin\alpha \cos\theta_\Omega \cos\phi}, \quad (A6)$$

$$\cos\theta_\mu = \cos\alpha \cos\theta_\Omega - \sin\alpha \sin\theta_\Omega \cos\phi, \quad (A7)$$

$$\sin\theta_\mu = \frac{\sin\alpha \sin\phi}{\sin\phi_\mu}, \quad (A8)$$

where $\chi = \theta_\Omega - \alpha$, $\alpha$ is the magnetic inclination between the magnetic axis and the axis of rotation, $\theta_\Omega$ is the angle between the axis of rotation and the *Z*-axis, and $\chi$ is called the impact angle, which represents the angle at which the angle between the LOS and the magnetic axis is smallest. When the magnetic axis is located in the *XZ* plane, $\theta_\mu$ equals $\chi$. We choose the radiation height $r_{em} = 50 R_s$ and assume that the polarization of the photon is in the $\boldsymbol{k}$–$\boldsymbol{B}$ plane (i.e., O-mode) and along the tangential direction of the local magnetic lines. It requires that the emission point $\boldsymbol{r}_i = (r_{em}, \theta_{ri}, \phi_{ri})$ satisfies

$$\theta_{ri} = \frac{\theta_{\mu i}}{2} - \frac{1}{2}\sin^{-1}\left(\frac{1}{3}\sin\theta_{\mu i}\right), \quad \phi_{ri} = \phi_{\mu i}. \quad (A9)$$

For a given radiation height $r_{em}$, and the initial rotation phase of the pulsar $\phi_i$, we can integrate the wave mode evolution equation along the path of photon propagation to obtain the final polarization state of the photon (Wang et al. 2010):

$$i\frac{d}{ds}\begin{pmatrix}A_+\\A_-\end{pmatrix} = \begin{bmatrix}-\Delta k/2 + \phi'_B \sin 2\theta_m & i(\theta'_m + \phi'_B \cos 2\theta_m)\\-i(\theta'_m + \phi'_B \cos 2\theta_m) & \Delta k/2 - \phi'_B \sin 2\theta_m\end{bmatrix}$$
$$\times \begin{pmatrix}A_+\\A_-\end{pmatrix},$$
$$(A10)$$

where the superscript (') denotes $d/ds$, $\Delta k = k_+ - k_- = \Delta n \omega/c$, and $\Delta n$ is the difference between the refractive indices of the two natural modes. $A_+$ and $A_-$ are the amplitudes of the natural wave mode. $\phi_B$ is the phase of the magnetic field in the *XYZ* system with $\tan\phi_B = B_Y/B_X$, and $\theta_m$ denotes the polarization of the natural mode, which is circularly polarized as it equals $\pi/4$, and linearly polarized as it equals 0 or $\pi/2$. In the *XYZ* coordinate system, we can give the polarization amplitude of the incident radiation by

$$\boldsymbol{E} = \begin{pmatrix} A_X \\ A_Y \end{pmatrix} = A_+ \boldsymbol{E}_+ + A_- \boldsymbol{E}_-, \quad (A11)$$

where

$$\boldsymbol{E}_+ = \begin{pmatrix} i\cos\theta_m \cos\phi_B - \sin\theta_m \sin\phi_B \\ i\cos\theta_m \sin\phi_B + \sin\theta_m \cos\phi_B \end{pmatrix},$$

$$\boldsymbol{E}_- = \begin{pmatrix} -i\sin\theta_m \cos\phi_B - \cos\theta_m \sin\phi_B \\ -i\sin\theta_m \sin\phi_B + \cos\theta_m \cos\phi_B \end{pmatrix}, \quad (A12)$$

and then the Stokes parameters can be given by

$$\begin{aligned} I &= A_X A_X^* + A_Y A_Y^* = |A_X|^2 + |A_Y|^2, \\ Q &= A_X A_X^* - A_Y A_Y^* = |A_X|^2 - |A_Y|^2, \\ U &= A_X A_Y^* + A_Y A_X^* \\ &= 2(\Re(A_X)\Re(A_Y) + \Im(A_X)\Im(A_Y)), \\ V &= -i(A_X A_Y^* - A_Y A_X^*) \\ &= -2(\Re(A_X)\Im(A_Y) - \Im(A_X)\Re(A_Y)), \end{aligned} \quad (A13)$$

where $\Re(f)$ and $\Im(f)$ denote the real and imaginary parts of $f$, respectively. Due to the rotation of the neutron star, we can only see the radiated photons when the cone angle at which the magnetic axis sweeps across the direction of the LOS (i.e., $Z$-axis), which corresponds to the change of the initial rotation phase $\phi_i$. The final polarization state will change with $\phi_i$, which is called the polarization profile. Furthermore, we consider the initial intensity distribution $I_0$ given by a Gaussian centered at $\phi_i = 0$, which gives us

$$I_0(\phi_i) = \exp(-4\sqrt{\ln 2}\, \phi_i^2 / \phi_{\max}^2), \quad (A14)$$

where $\phi_{\max}$ is the initial phase of the photon when it is at the edge of the open field region and can be written as

$$\cos\phi_{\max} = \frac{\cos\theta_{\text{open}} - \cos\theta_\Omega \cos\alpha}{\sin\theta_\Omega \sin\alpha}, \quad (A15)$$

where $\theta_{\text{open}} = \sqrt{r_{\text{em}}/r_{\text{lc}}}$ is the half opening angle at height $r_{\text{em}}$. We consider the typical parameters: the positive and negative electrons move with the same single Lorentz factor $\gamma = 100$, the multiplicity $\eta = 400$, the period of the neutron star $P = 1$ s, and $B_s = 10^{12}$ G.

## ORCID iDs

Ze-Nan Liu https://orcid.org/0000-0002-5758-1374
Shu-Qing Zhong https://orcid.org/0000-0002-1766-6947
Fa-Yin Wang https://orcid.org/0000-0003-4157-7714
Zi-Gao Dai https://orcid.org/0000-0002-7835-8585

## References

Backer, D. C. 1976, ApJ, 209, 895
Beniamini, P., Kumar, P., & Narayan, R. 2022, MNRAS, 510, 4654
Blandford, R. D., & Scharlemann, E. T. 1976, MNRAS, 174, 59
Bochenek, C. D., Ravi, V., Belov, K. V., et al. 2020, Natur, 587, 59
Caleb, M., Keane, E. F., van Straten, W., et al. 2018, MNRAS, 478, 2046
Caleb, M., Stappers, B. W., Abbott, T. D., et al. 2020, MNRAS, 496, 4565
Camilo, F., Ransom, S. M., Halpern, J. P., et al. 2006, Natur, 442, 892
Chen, B. H., Hashimoto, T., Goto, T., et al. 2022, MNRAS, 509, 1227
Chen, H.-Y., Gu, W.-M., Sun, M., & Yi, T. 2022, ApJ, 939, 27
Cheng, A. F., & Ruderman, M. A. 1979, ApJ, 229, 348
CHIME/FRB Collaboration, Andersen, B. C., Bandura, K., et al. 2019, ApJL, 885, L24
CHIME/FRB Collaboration, Andersen, B. C., Bandura, K. M., et al. 2020, Natur, 587, 54
Cho, H., Macquart, J.-P., Shannon, R. M., et al. 2020, ApJL, 891, L38
Connor, L., Miller, M. C., & Gardenier, D. W. 2020, MNRAS, 497, 3076
Dai, S., Hobbs, G., Manchester, R. N., et al. 2015, MNRAS, 449, 3223
Dai, S., Lower, M. E., Bailes, M., et al. 2019, ApJL, 874, L14
Dai, S., Lu, J., Wang, C., et al. 2021, ApJ, 920, 46
Day, C. K., Deller, A. T., Shannon, R. M., et al. 2020, MNRAS, 497, 3335
Feng, Y., Li, D., Yang, Y.-P., et al. 2022a, Sci, 375, 1266
Feng, Y., Zhang, Y.-K., Li, D., et al. 2022b, SciBu, 67, 2398
Gajjar, V., Siemion, A. P. V., Price, D. C., et al. 2018, ApJ, 863, 2
Goldreich, P., & Julian, W. H. 1969, ApJ, 157, 869
Hashimoto, T., Goto, T., On, A. Y. L., et al. 2020a, MNRAS, 498, 3927
Hashimoto, T., Goto, T., Wang, T.-W., et al. 2020b, MNRAS, 494, 2886
Hessels, J. W. T., Spitler, L. G., Seymour, A. D., et al. 2019, ApJL, 876, L23
Ikebe, S., Takefuji, K., Terasawa, T., et al. 2023, PASJ, 75, 199
Jackson, T., Goto, T., Wang, T.-W., et al. 1998, MNRAS, 494, 2886
Jiang, J.-C., Wang, W.-Y., Xu, H., et al. 2022, RAA, 22, 124003
Josephy, A., Chawla, P., Fonseca, E., et al. 2019, ApJL, 882, L18
Kim, S. J., Hashimoto, T., Chen, B. H., et al. 2022, MNRAS, 514, 5987
Kirsten, F., Snelders, M. P., Jenkins, M., et al. 2021, NatAs, 5, 414
Kumar, P., Lu, W., & Bhattacharya, M. 2017, MNRAS, 468, 2726
Liu, Z.-N., Geng, J.-J., Yang, Y.-P., Wang, W.-Y., & Dai, Z.-G. 2023, ApJ, 958, 35
Liu, Z.-N., Wang, W.-Y., Yang, Y.-P., & Dai, Z.-G. 2023b, ApJ, 943, 47
Lorimer, D. R., Bailes, M., McLaughlin, M. A., Narkevic, D. J., & Crawford, F. 2007, Sci, 318, 777
Lorimer, D. R., & Kramer, M. 2012, Handbook of Pulsar Astronomy (Cambridge: Cambridge Univ. Press)
Luo, J.-W., Zhu-Ge, J.-M., & Zhang, B. 2023, MNRAS, 518, 1629
Luo, R., Wang, B. J., Men, Y. P., et al. 2020, Natur, 586, 693
Lyne, A., Levin, L., Stappers, B., et al. 2018, ATel, 12284, 1
Masui, K., Lin, H.-H., Sievers, J., et al. 2015, Natur, 528, 523
Metzger, B. D., Margalit, B., & Sironi, L. 2019, MNRAS, 485, 4091
Michilli, D., Seymour, A., Hessels, J. W. T., et al. 2018, Natur, 553, 182
Nimmo, K., Hessels, J. W. T., Keimpema, A., et al. 2021, NatAs, 5, 594
Petroff, E., Barr, E. D., Jameson, A., et al. 2016, PASA, 33, e045
Petroff, E., & Yaron, O. 2020, TNSAN, 160, 1
Platts, E., Caleb, M., Stappers, B. W., et al. 2021, MNRAS, 505, 3041
Pleunis, Z., Good, D. C., Kaspi, V. M., et al. 2021, ApJ, 923, 1
Prochaska, J. X., Macquart, J.-P., McQuinn, M., et al. 2019, Sci, 366, 231
Qu, Y., Kumar, P., & Zhang, B. 2022, MNRAS, 515, 2020
Qu, Y., & Zhang, B. 2023, MNRAS, 522, 2448
Qu, Y., Zhang, B., & Kumar, P. 2023, MNRAS, 518, 66
Radhakrishnan, V., & Cooke, D. J. 1969, ApL, 3, 225
Radhakrishnan, V., & Rankin, J. M. 1990, ApJ, 352, 258
Ravi, V., Shannon, R. M., Bailes, M., et al. 2016, Sci, 354, 1249
Tong, H., Wang, P. F., Wang, H. G., & Yan, Z. 2021, MNRAS, 502, 1549
Wang, C., & Lai, D. 2009, MNRAS, 398, 515
Wang, C., Lai, D., & Han, J. 2010, MNRAS, 403, 569
Wang, F. Y., Zhang, G. Q., Dai, Z. G., & Cheng, K. S. 2022, NatCo, 13, 4382
Wang, W., Zhang, B., Chen, X., & Xu, R. 2019, ApJL, 876, L15
Wang, W.-Y., Jiang, J.-C., Lu, J., et al. 2022a, SCPMA, 65, 289511
Wang, W.-Y., Xu, R., & Chen, X. 2020, ApJ, 899, 109
Wang, W.-Y., Yang, Y.-P., Niu, C.-H., Xu, R., & Zhang, B. 2022b, ApJ, 927, 105
Xiao, D., & Dai, Z.-G. 2022, A&A, 667, A26
Xu, J., Feng, Y., Li, D., et al. 2023, Univ, 9, 330
Yang, Y.-P. 2023, ApJ, 956, 67
Yang, Y.-P., Lu, W., Feng, Y., Zhang, B., & Li, D. 2022, ApJL, 928, L16
Yang, Y.-P., & Zhang, B. 2018, ApJ, 868, 31
Yang, Y.-P., Zhu, J.-P., Zhang, B., & Wu, X.-F. 2020, ApJL, 901, L13
Zhang, B. 2018, The Physics of Gamma-Ray Bursts (Cambridge: Cambridge Univ. Press)
Zhang, B. 2022, ApJ, 925, 53
Zhang, Y.-K., Li, D., Zhang, B., et al. 2023, ApJ, 955, 142
Zhong, S.-Q., Xie, W.-J., Deng, C.-M., et al. 2022, ApJ, 926, 206
Zhu, W., Xu, H., Zhou, D., et al. 2023, SciA, 9, eadf6198
Zhu-Ge, J.-M., Luo, J.-W., & Zhang, B. 2023, MNRAS, 519, 1823